\documentclass[aps,prd,email,preprint,showpacs,showkeys,preprintnumbers,amsmath,amssymb,nofootinbib]{revtex4}

\usepackage{color}
\usepackage{latexsym}
\usepackage{amsmath}
\usepackage{amssymb}
\usepackage{eufrak}
\usepackage{euscript}
\usepackage{pstricks}
\usepackage{graphics}
\usepackage{graphicx}
\usepackage{picture}
\def\[{\left\lbrack}
\def\]{\right\rbrack}

\def\({\left(}
\def\){\right)}

\newcommand{\bbe}{\begin{equation}}
\newcommand{\eee}{\end{equation}}
\newcommand{\eaa}{\end{eqnarray}}
\newcommand{\baa}{\begin{eqnarray}}

\def\ni{\noindent}

\begin{document}

\pagestyle{myheadings}
\markright{Tsallis' thermostatitics, MOND theory and holographic......}

\title{\Large{Tsallis' thermostatitics, MOND theory and \\ holographic considerations on a Machian Universe}}

\author{Everton M. C. Abreu$^{a,b}$}
\email{evertonabreu@ufrrj.br}
\author{Jorge Ananias Neto$^{b}$}
\email{jorge@fisica.ufjf.br}

\affiliation{$^a$Grupo de F\' isica Te\'orica e Matem\'atica F\' isica, Departamento de F\'{\i}sica, \\
Universidade Federal Rural do Rio de Janeiro\\
BR 465-07, 23890-971, Serop\'edica, RJ, Brazil \\\\ 
$^b$Departamento de F\'{\i}sica, ICE, Universidade Federal de Juiz de Fora,\\
36036-330, Juiz de Fora, MG, Brazil\\\\
\today\\}
%\emailAdd{evertonabreu@ufrrj.br, jorge@fisica.ufjf.br, \\ albert@fisica.ufjf.br, wilson@fisica.ufjf.br}

%\arxivnumber{....arxiv:}

\begin{abstract}
\noindent MOND theory explains the rotation curves of the galaxies.  Verlinde's ideas establishes an entropic origin for gravitational forces and Tsallis principle generalizes the theory of Boltzman-Gibbs.  In this work we have promoted a connection between these recent approaches, that at first sight seemed to have few or no points in common, using the Mach's principle as the background.  In this way we have used Tsallis formalism to calculate the main parameters of the Machian Universe including the Hubble parameter and the age of the Universe.  After that, we have also obtained a new value for the Tsallis parameter via Mach's principle.  Using Verlinde's entropic gravity we have obtained new forms for MOND's well established ingredients.   Finally, based on the relations between particles and bits obtained here, we have discussed the idea of bits entanglement in the holographic screen.
\end{abstract}

% \PACS {04.50.-h, 05.20.-y, 05.90.+m}

\maketitle

\pagestyle{myheadings}
\markright{Tsallis' thermostatitics, MOND theory and holographic......}

%%%%%%%%%%%%%%%%%%%%%%%%%%%%%%%

%%%%%%%%%%%%%%%%%%%%%%%%%%%%%%%
%\newcommand{\be}{\begin{equation}}
%\newcommand{\ee}{\end{equation}}
%\newcommand{\ba}{\begin{eqnarray}}
%\newcommand{\ea}{\end{eqnarray}}
%\newcommand{\p}{\partial}
%\def\ni{\noindent} 
%\def\Dc{{\cal D}}
%\def\no{\nonumber}

%%%%%%%%%%%%%%%%%%%%%%%%%%%%%%%%%%%%%%%%%%%%%%%%%%%%%%%%%%%%%%%%%%%%%%%%%%%%%%%%%%%%%%%%%%%%%%%%%%%%%%%%%%%%%%%%%%%%%%%%%%%

%\renewcommand{\theequation}{1.\arabic{equation}}
%\setcounter{equation}{0}

%\newpage

\section{Introduction}
\renewcommand{\theequation}{1.\arabic{equation}}
\setcounter{equation}{0}

One of the great mysteries of general physics at current times is to explain dark matter, which composes the great majority of the matter of the Universe.  Therefore, there are several reasons to derive alternatives to dark matter, following \cite{BEK2} we will name three good reasons.  The first one is a kind of alternative just in case our understanding of gravity on astronomical scales, in which the concept of dark matter is based, is not correct.  The second one would be based on the fact that dark matter has not been identified directly yet.  In other words, it was not identified through gravitational experiments.  And the last reason, the simpler one, is that it is always good to have other paradigms of it in order to be able to confront them, since the dark matter paradigm is not fully understood yet.

Besides, another challenge comes from the well known fact that classical Newtonian dynamics does not work on galactic scales.  Measures taken through observations on the rotation curves showed that galaxies are not rotating in the same way as the Solar system.  The velocity of rotation versus the distance from the galactic center are visualized through the so-called galactic rotation curves.  However, the last ones cannot be understood through only the visible matter.  This fact lead us to consider the possibility, among other, that the Newtonian dynamics is not valid universally speaking.

One explanation came with Zwicky \cite{Zwicky, Zwicky2} that considered dark matter, in addition to the matter of the known galaxies, taking part in the galactic scale by its gravitational action.  However, nowadays we know that dark matter is not only an exotic feature of clusters.   It can also be found in single galaxies to explain their flat rotation curves.

To explore this problem, Milgrom \cite{Milgrom, Milgrom2, Milgrom3} proposed a second explanation which results in the so-called modified Newtonian dynamics, MOND, as a phenomenological theory.  The basic ingredient in MOND is to modify the force expression to adapt it to the case of extremely small accelerations, that occurs at the scale of galaxies.

Concerning the force expression and gravitation we know that recently E. Verlinde \cite{Verlinde} brought an idea that the gravitational force appears as the result of entropic and holographic concepts.  Through these first principles Verlinde was able to obtain Newton's gravitation law but he did not deal with extremely small accelerations.

The holographic screen, formed by bits of informations, on the other hand can be considered as the causal sphere that encompasses the two-body interaction and all other matter that affects this two-body, which agrees with Mach principle.  Concerning Mach's ideas \cite{mach,mach3}, in the scenario of gravitational interaction of closed objects, one can substitute the far Universe by a spherical shell of effective mass $M$ and radius $R$.  This spherical shell, in our point of view, can be seen as the holographic screen.  Hence, the mass inside, according to Mach's principle, has a constant gravitational potential which will be described latter.

It is our objective in this work to analyze the connection between Mach's and Verlinde's ideas in the light of MOND formalism.  Moreover, we derived Mach's underlying ingredients as functions of Tsallis' nonextensive parameter \cite{Tsallis,Tsallis2}.
                                                                                                                                                                                                
The organization of this paper follows a structure where the section II and III are dedicated to introductory explanations
of Verlinde's and MOND formalisms, respectively.  In section IV and V, after a very brief introduction of the important ingredients of Mach and Tsallis principles that will be used here, respectively, we will calculate the value of the MOND parameters through Tsallis parameter in section VI.  After that, in section VII we will make a connection between Verlinde's bits-holographic-screen ideas and Mach's particles-horizon concepts.   In section VIII we will introduce new values for some MOND ingredients in a Machian model.  In section IX we will discuss the results and conclusions obtained in this the work.

\section{A Brief Review of Verlinde's Formalism}
\label{vf}
\renewcommand{\theequation}{2.\arabic{equation}}
\setcounter{equation}{0}

The objective of this section is to provide the reader with the main tools that will be used in the following sections.   Although both formalisms are well known in the literature, these brief reviews can emphasize precisely that there is a connection between both ideas and that it was established recently \cite{ananias,aa,aa2}.

The study of entropy has been an interesting task through recent years thanks to the fact that it can be understood as a measure of information loss concerning the microscopic degrees of freedom of a physical system, when describing it in terms of macroscopic variables.  Appearing in different scenarios, we can conclude that entropy can be considered as a consequence of the gravitational framework \cite{BEK,HAW}.

The formalism proposed by E. Verlinde \cite{Verlinde} obtains the gravitational acceleration  by using the holographic principle and the well known equipartition law of energy. His ideas relied on the fact that gravitation can be considered universal and independent of the details of the spacetime microstructure \cite{nicolini}.  Besides, he brought new concepts concerning holography since the holographic principle must unify matter, gravity and quantum mechanics.

The model considers a spherical surface as being the holographic screen, with a particle of mass $M$ positioned in its center. The holographic screen can be imagined as a storage device for information. The number of bits, which is the smallest unit of information in the holographic screen, is assumed to be proportional to the  holographic screen
area $A$
\begin{eqnarray}
\label{bits}
N = \frac{A }{l_P^2},
\end{eqnarray}
where $ A = 4 \pi r^2 $ and $l_P = \sqrt{\frac{G\hbar}{c^3}}$ is the Planck length and $l_P^2$ is the Planck area.   In Verlinde's framework one can suppose that the bits total energy on the screen is given by the equipartition law of energy

\begin{eqnarray}
\label{eq}
E = \frac{1}{2}\,N k_B T.
\end{eqnarray}
It is important to notice that the usual equipartition theorem in Eq. (\ref{eq}), can be derived from the usual Boltzmann-Gibbs thermostatistics. 
%We will see that in a nonextensive thermostatistics scenario, the equipartition law of energy will be modified in a sense that a nonextensive parameter $q$ will be introduced in its expression.
Let us consider that the energy of the particle inside the holographic screen is equally divided by all bits in such a manner that we can have the expression

\begin{eqnarray}
\label{meq}
M c^2 = \frac{1}{2}\,N k_B T.
\end{eqnarray}
With Eq. (\ref{bits}) and using the Unruh temperature equation \cite{unruh} given by

\begin{eqnarray}
\label{un}
k_B T = \frac{1}{2\pi}\, \frac{\hbar a}{c},
\end{eqnarray}
we are  able to obtain the  (absolute) gravitational acceleration formula

\begin{eqnarray}
\label{acc}
a &=&  \frac{l_P^2 c^3}{\hbar} \, \frac{ M}{r^2}\nonumber\\ 
&=& G \, \frac{ M}{r^2}\,\,.
\end{eqnarray}
From Eq. (\ref{acc}) we can see that the Newton constant $G$ is just written in terms of the fundamental constants, $G=l_P^2 c^3/\hbar$.

\section{The MOND Theory and Modified Entropic Force}
%\label{TTS}
\renewcommand{\theequation}{3.\arabic{equation}}
\setcounter{equation}{0}
%%%%%%%%%%%%%%%%%%%%%%%%%%%%%%%%%%%%%%%%%%%%%%%%%%%%%%%%%%%%%%%%%%%%%%%%%%%%%%%%%%%%%%%%%%%%%%%%%%%%%%%%%%%%%%%%%%%%%%%%%%%

Phenomenologically speaking, MOND theory is able, in principle, to explain the majority of the galaxies rotation curves. Besides, MOND theory can recover the well known Tully-Fisher relation \cite{TF}.   It can also be an alternative to the dark matter model. However, it is important to mention that MOND theory is not able to explain the temperature scenario of galaxy clusters and it shows some problems when comparing with cosmology. For details, see for instance, Refs. \cite{Sk} and \cite{NP}.
To sum up, this theory is a modification of Newton's second law in which the force can be depicted as
\bbe \label{3.1}
F=m\; \mu\(\frac{a}{a_0}\)\,a\,\,,
\eee
where $\mu=\mu(x)$ is a function with the following properties: $\mu(x)\approx 1$ for $ x > > 1$, $\mu(x) \approx x$ for $ x < < 1$.
The $ a_0 $ parameter is a constant.
There are different interpolation functions for $\mu(x)$ (see \cite{form}). However, we can believe that the main consequences caused by MOND theory  do not rely on the specific form of this function. Therefore, for simplicity, it is usual to suppose that the variation of $\mu(x)$ between the asymptotic limits occurs abruptly at $x=1$ or $a=a_0$.

In order to derive MOND's theory from Verlinde's formalism, one of us \cite{ananias} have considered that, below a critical temperature, the cooling down of the holographic screen is not homogeneous. We will choose that the fraction of bits with zero energy will be given by the formula
\begin{eqnarray}
\label{n0}
\frac{N_0}{N} = 1-\frac{T}{T_c},
\end{eqnarray}
where $N$ is the total number of bits, $N_0$ is the number of bits with zero energy and $T_c$ is the critical temperature.
For $ T \geq T_c $ we have that $N_0=0$ and for $T < T_c$ the zero energy phenomenon for some bits begins to appear. 
Eq. (\ref{n0}) is an standard relation of critical phenomena and second order phase transitions theory. The number of bits with energy different from zero for a given temperature $ T < T_c \,$ is

\begin{eqnarray}
\label{sub}
N-N_0 = N \frac{T}{T_c}.
\end{eqnarray}
Let us assume that Eq. (\ref{bits}) is still true below $T_c$. We know that the energy of the particle inside the holographic screen is equally distributed by all bits with nonzero energy together with the equipartition law of energy, so

\begin{eqnarray}
\label{ebec}
M c^2 &=& \frac{1}{2} (N-N_0) k_B T \nonumber\\\nonumber\\ 
&=& \frac{1}{2}\, N \frac{T}{T_c}\; k_B T.
\end{eqnarray}
Then we are able to derive,  for $ T < T_c $, the MOND theory for Newton's law of gravitation
\begin{eqnarray}
\label{amond}
a\,\(\frac{a}{a_0}\)=G \frac{M}{r^2},
\end{eqnarray} 
where 
\begin{eqnarray}
\label{a0}
a_0=\frac{2\pi c\, k_B  T_c}{\hbar}. 
\end{eqnarray}
Using that $a_0\approx 10^{-10}\, m\,s^{-2}$ we obtain $T_c \approx 10^{-31} K$.  It is  an extremely low temperature which is far from the standard temperatures observed in our real world.  Here, a brief comment on this extremely low temperature must be made. Rewriting Eq. (\ref{un}) in the form
\begin{eqnarray}
T \approx 4 \times 10^{-21} \, \frac{K}{m/s^2}\; a\,\,,
\end{eqnarray}
we can notice that, at first sight, any model that combines the Unruh temperature with MOND theory ($a_0\approx 10^{-10}\, m\, s^{-2}$) leads to the critical temperature, $T_c$, mentioned above. The Unruh effect, where an accelerating observer will observe a black-body radiation, establishes that our critical temperature 
should be observed in an accelerated reference frame that can be the galaxies.
Therefore, below the critical temperature, $T_c$, we have obtained, as a result of MOND theory, that the standard Newton's second law is no longer valid. 

The hypothesis of a nonhomogeneous bits cooling down can be justified by making use of thermostatistic arguments. Below the critical (low) temperature, the thermal bath does not guarantee an exact application of the equipartition theorem for all bits.  Namely, the system thermalization does not guarantee that all bits have the same energy. This result frequently happens in several different physical systems. As an example, we can mention the well known coupled harmonic oscillators model \cite{chom} where the energy is not distributed between the independent normal modes of this system.

\section{The Machian Universe}
\renewcommand{\theequation}{4.\arabic{equation}}
\setcounter{equation}{0}
\label{Res}

In a Machian model of the Universe, all the particles in the Universe are non-locally interacting with each other.  The consequence is that the Universe can be deemed as a statistical ensemble of gravitationally entangled particles \cite{darabi,ldhd,gogberashvili}.

E. Mach, in his work about the science of mechanics, analyzed the physical foundations of Newtonian mechanics \cite{mach3}.  He suggested that inertia is a consequence of the mutual interaction between masses.

Namely, when we consider the gravitational interaction of close objects, we can substitute the distant Universe by a spherical shell of effective mass $M$ and the effective radius $R$ which involves the ensemble of $N_p$ (the number of particles from now on) uniformly distributed identical particles which have gravitational mass $mg$ each one.  The non-local Machian interaction originates the so-called classical spacetime.  From the Machian model one can assume that the gravitational potential of any particle from the ensemble can be written as
\bbe
\label{4.1}
\Phi\,=\,-\frac{2 M_U G}{R}\,\,,
\eee
where the potential of the whole Universe ($\Phi$) originates the speed of light.   $M_U$ is the total mass of the Universe, which has a radius $R$.  $G$ is the Newton constant.  We can see clearly from (\ref{4.1}) that this universal potential $\Phi$ and consequently $c$ can be considered as constants since the Universe is homogeneous and isotropic on the horizon scales $R$ \cite{gogberashvili}.  For each particle in the gravitationally entangled Universe interacts with all the other $(N_p -1)$ particles.  Hence, we have that $N_p (N_p -1)/2$ is the number of interacting pairs.  The Machian energy of a single particle is given  by

\begin{eqnarray}
\label{4.2}
E_0\,&=&\,\frac{N_p (N_p -1)}{2}\,\frac{2 G m^2_g}{R}\,\nonumber\\
&\approx& \,-\frac{G N^2_p m^2_g}{R}\,=\,-m_g \Phi\,\,,
\end{eqnarray}
and it can be shown that the observed gravity strength weakens by a factor related to the number of particles in the Universe.

It is interesting to notice that Eq. (\ref{4.1}) is equivalent to the critical density condition in relativistic cosmology, namely
\bbe \label{4.3}
\rho_c\,=\,\frac{3 M_U}{4 \pi R^3}\,=\,\frac{3 H^2}{8 \pi G}\,\,,
\eee
where $H\sim\,c/R$ is the well known Hubble constant.

Using Eq. (4.3), we are now in condition to compute (approximately) the total mass of the Universe, which is
\bbe \label{4.4}
M_U \sim \frac{c^3}{2 G H} \approx 10^{53} kg\,\,,
\eee
and Eq. (\ref{4.1}) leads us to construct the Mach principle which relates the origin of inertia of a particle, or its rest energy, to particle's interactions concerning the whole Universe, as
\bbe \label{4.5}
E\,=\,m c^2\,=\,-m \Phi\,\,,
\eee
where $m$ describes the inertia of the particle which is not always constant.  In general it is not constant.  In (\ref{4.5}), the universal Machian potential considers the contribution of the collective gravitational interactions between all $N_p$ particles inside the horizon.  As we said before, since each particle interacts with all other $(N_p -1)$ particles, we have that the total Machian energy comprises $N_p (N_p -1)/2$ terms of magnitude $\approx\, 2Gm^2 /R$ (the mean separation between interacting pairs is $R/2$ \cite{gogberashvili}).    For very large $N_p$, the Machian energy of a single interacting particle with the total Machian potential $\Phi$, can be written as
\bbe \label{4.6}
E \approx N_p^2 \frac{G m^2}{R}\,\,.
\eee
We can now write the contribution of the collective Machian interactions to the total mass of the Universe, which is
\bbe \label{4.7}
M_{Mach} \approx \frac 12 N^2_p m\,\,,
\eee
and we can say that the total mass of the Universe is given by
\bbe \label{4.7.1}
M_U \sim M_{Mach}\,\,,
\eee
which is of the order $N^2_p$ \cite{gogberashvili} and not $N_p m$.

The response time of the Universe that ``feels" the motion of a quantum particle can be approximately calculated as being
\bbe \label{4.8}
\Delta t \sim \frac{R}{N_p c} \sim \frac{1}{N_p H}\,\,,
\eee
where $\Delta t$ results from a delayed response of the whole ensemble.  Consequently, any mechanical process in the world ensemble carries an exchange of at least the minimal quantity of the action
\bbe \label{4.8.1}
A\,=\,m c^2 \Delta t\,\,,
\eee
which can be identified with Planck's quantum action
\bbe \label{4.9}
A \,=\,- \int dt E \approx -\,m c^2 \Delta t \,=\, 2 \pi \hbar\,\,,
\eee
and from Eqs. (\ref{4.4}), (\ref{4.8}) and (\ref{4.9}) we can compute the total action of the Universe as being
\bbe \label{4.10}
A_U\,=\,\frac{M_U c^2}{H} \approx \frac{N_p^3}{2} A\,\,,
\eee
and the number of particles within is
\bbe \label{4.11}
N_p \approx \Big( \frac{2 A_U}{H} \Big)^{\frac 13} \approx \Big( \frac{M_U c^2}{\pi \hbar H} \Big)^{\frac 13} \approx 10^{40}\,\,,
\eee
which is one of the main parameters of the Machian model.  This number is considered to pinpoint the relation between micro and macro physics \cite{gogberashvili}.

Using Eqs. (\ref{4.11}), (\ref{4.7}) and (\ref{4.4}) one can calculate the mass of a particle in this description of a simplified Machian Universe, which is 
\bbe \label{4.12}
m \approx \frac{2 M_U}{n^2} \approx 2 \cdot 10^{-27} kg \approx 1 GeV c^{-2}\,\,,
\eee
which is closer to the proton mass.

Following Verlinde's point of view, in a Machian Universe, each particle carries its holographic surface formed by $N_b$ bits of information.   Since each particle is gravitationally entangled and form a statistical ensemble we will apply the nonextensive analysis to discuss this property.

\section{A Brief review of Tsallis' formalism}
\renewcommand{\theequation}{5.\arabic{equation}}
\setcounter{equation}{0}

An important formulation of the nonextensive (NE) Boltzmann-Gibbs thermostatistics has been proposed by Tsallis \cite{Tsallis, Tsallis2} in which the entropy is given by the formula

\begin{eqnarray}
\label{nes}
S_q =  k_B \, \frac{1 - \sum_{i=1}^W p_i^q}{q-1}\;\;\;\; (\sum_{i=1}^W p_i = 1)\,\,,
\end{eqnarray}
where $p_i$ is the probability of the system to be in a microstate, $W$ is the total number of configurations and $q$, known in the current literature as Tsallis parameter or NE parameter, is a real parameter quantifying the degree of nonextensivity. 
The definition of entropy (\ref{nes}) has as motivation to study multifractals systems and it also possesses the usual properties of positivity, equiprobability, concavity and irreversibility.
It is important to note that Tsallis' formalism contains the Boltzmann-Gibbs statistics as a particular case in the limit $ q \rightarrow 1$ where the usual additivity of entropy is recovered. Plastino and Lima \cite{PL}, by using a generalized velocity distribution for free particles given by

\begin{eqnarray}
f_0(v) = B_q \[ 1-(1-q) \frac{m v^2}{2 k_B T} \]^{1/1-q}\,\,,
\end{eqnarray}
where $B_q$ is a $q$-dependent normalization constant, $m$ and $v$ is a mass and velocity of the particle, respectively, have derived a NE equipartition law of energy whose expression can be written as,

\begin{eqnarray}
\label{ge}
E = \frac{1}{5 - 3 q} N k_B T\,\,,
\end{eqnarray}
where the range of $q$ is $ 0 \le q < 5/3 $.  For $ q=5/3$ (critical value) the expression of the equipartition law of energy, Eq. (\ref{ge}), diverges. It is easy to observe that for $q = 1$,  the classical equipartition theorem for each microscopic degrees of freedom is recovered.

\section{Nonextensive considerations about Machian principle and a new value for $q$}
\renewcommand{\theequation}{6.\arabic{equation}}
\setcounter{equation}{0}

In this section we will promote a connection between the main ingredients of the Mach model and, considering that the Mach's particles constitute a statistical ensemble, the NE ideas of Tsallis statistical mechanics.   Having said that, let us begin with the relation defined 
in \cite{ananias}, given by

\begin{eqnarray}
\label{a1}
G_{NE} = \frac{5 - 3 q}{2} \,G\,\,,
\end{eqnarray}
and introduce it into the relation concerning the universal potential $\Phi$ (Eq. (4.1)) as

\begin{eqnarray}
\label{a2}
\Phi&=&-\frac{2 M_U G_{NE}}{R}\,\,, \nonumber \\
&=&-\,\frac{5-3q}{R}\,M_U\, G
\end{eqnarray} 
which, as we know, it is equal to $-c^2$.   Hence, from Eq. (\ref{a2}) we can write that

\begin{eqnarray}
\label{a3}
\Phi= - \frac{5-3q}{R}\, M_U G =-c^2\,\,,
\end{eqnarray}

\ni and consequently the Universe radius is given by

\begin{eqnarray}
\label{a4}
R\,=\,\frac{5-3q}{c^2}\,M_U\,G\,\,,
\end{eqnarray}

\ni which allows us to calculate the response time in Eq. (4.9) as a function of the NE parameter $q$ as

\begin{eqnarray}
\label{a5}
\Delta t \,\sim \, \frac{5-3q}{N_p c^3} \,M_U\,G
\end{eqnarray}

\ni and this result for $\Delta t$ shows us that the well known bound for $q$, i.e., $q \leq 5/3$ is preserved, since $\Delta t \geq 0$.  Since when $q=5/3$ we have that $\Delta t=0$ which is a non-causal result for $\Delta t$.  Hence, we have that $q < 5/3$.

We can also calculate the Hubble constant, $H \sim c/R$, as a function of the NE parameter $q$.   Hence, using Eq. (\ref{a5}) we have that,

\begin{eqnarray}
\label{a6}
H\,\sim\,\frac{1}{5-3q}\,\frac{c^3}{M_U\,G}
\end{eqnarray}

\ni which can provide an expression for the scale factor.

Another important parameter is the critical density condition, Eq. (4.3), which will be given by 

\begin{eqnarray}
\label{a7}
\rho_c=\frac{3 H^2}{8\pi G_{NE}}=\frac{6 H^2}{(5-3q) G}.
\end{eqnarray}

\ni and using Eq. (\ref{a6}) we have that 

\begin{eqnarray}
\label{a8}
\rho_c\,=\,\frac{6 c^6}{(5-3q)^3,M_U G}
\end{eqnarray}

\ni and since $\rho_c > 0$ we have again that $q < 5/3$.

Let us analyze Eq. (\ref{a5}) concerning the response time.  Following reference [25] its estimation is given by

\begin{eqnarray}
\label{a9}
\Delta t \,\approx\,0.5 \cdot 10^{-22}\,s
\end{eqnarray}

\ni and the number of typical particles in the Universe is [25]

\begin{eqnarray}
\label{a10}
N_p\,\approx\,10^{40}
\end{eqnarray}

Substituting Eqs. (\ref{a9}) and (\ref{a10}) into Eq. (\ref{a5}) we have that, for the NE parameter

\begin{eqnarray}
\label{a11}
5\,-\,3q \,\sim\,\frac{\Delta t N_p c^3}{M_U G}
\end{eqnarray}

\ni and using that $M_U \approx 10^{53} kg$ [25] the value is

\begin{eqnarray}
\label{a12}
5-3q\,=\,2.16
\end{eqnarray}
\begin{eqnarray}
\label{a13}
\Longrightarrow \quad 1\,-\,q\,\approx 0.053
\end{eqnarray}
\begin{eqnarray}
\label{a14}
\Longrightarrow \quad q\,\approx\,0.9467\,\,,
\end{eqnarray}

\ni which establish a new value for $q$.  Notice that concerning Eq. (5.3) we have analyzed that for $q \rightarrow 1$ we have that $5-3q = 2$ and the classical equipartition theorem is recovered.  Besides, with the result in Eq. (\ref{a12}) using only the Machian parameters, we have obtained a value very close to the established one, i. e., 2.16 and $q \approx 0.9467$ very close to 1.

Using the same values to calculate the Hubble parameter in Eq. (\ref{a6}) we have that

\begin{eqnarray}
\label{a15}
H\,\sim\, 12.5 \cdot 10^{-18}\,s^{-1}
\end{eqnarray}

\ni and for the age of the Universe we can write 

\begin{eqnarray}
\label{a16}
T_U \,\approx \, \frac{R}{c}\,&=&\,\frac{5-3q}{c^3}\,M_U G \nonumber \\
&=&3.1536 \cdot 10^7 \,s \nonumber \\
&=&16.92 \cdot 10^9\,\mbox{years}
\end{eqnarray}

\ni which is very close to the well known current estimation.

The numerical results obtained in Eqs. (\ref{a12}) - (\ref{a16}) are very closed to the ones obtained in the literature.  We believe that these almost equal results show us that the substitution $G=G_{NE}$ made in Eq. (\ref{a2}) is correct, which demonstrate the statistical nature of the Machian model.

In the next section we will try to establish a connection between Verlinde's formalism and the Mach principle.  This attempt was motivated by the holographic principle and the long range interactions considered in the Machian model.

\section{Bits entanglement}
\renewcommand{\theequation}{7.\arabic{equation}}
\setcounter{equation}{0}

Let us begin considering the NE equipartition law of energy applied to a particle inside the holographic screen, i.e., Eqs (2.3) and (5.3)

\begin{eqnarray}
\label{aA}
m c^2\,=\,\frac{1}{5-3q}\,N_b\,k_B\,T\,\,,
\end{eqnarray}

\ni where $N_b$ is the number of bits (from now on) distributed in the holographic screen.

From Mach ideas, Eq. (4.13) we can see that the number of particles is given by

\begin{eqnarray}
\label{aB}
N_p^3\,=\,\frac{M_U c^2}{\pi \hbar H}
\end{eqnarray}

\ni and from equations (\ref{aA}) and (\ref{aB}) we can obtain that

\begin{eqnarray}
\label{aC}
N_b\,=\,\kappa N^3_p\,\,,
\end{eqnarray}

\ni where $$\kappa = \frac{m\pi \hbar H}{M_U k_B T}$$ and Eq. (\ref{aC}) shows us a simple relation between the number of bits of the holographic screen and the number of particles inside it.  It is natural that the number of bits be greater than the number of particles.  Moreover, Eq. (\ref{aC}) demonstrate that the connection between Verlinde's and Mach's formalism is direct.

Following this path. since Eq. (\ref{aB}) has a quantum feature we can ask if it can pinpoint the quantum feature of an entanglement projection.   In other words, since Eq. (\ref{aC}) shows a quantum behavior, we can think that it may be a manifestation of the entanglement that must exist between the pair bit-bit within the holographic screen.   Namely, the bit-bit entanglement would be the projection of the long range interactions, described in Machian model between the pairs particle-particle.  The horizon defined in the Mach principle would be the a kind of mapping of the holographic screen. Namely, the bits entanglement in the holographic screen is the quantum correspondence, because entanglement is a quantum feature described by quantum field theory, of the interactions between the particles inside the holographic screen.  We can ask if this connection between particles' long range interactions described by the Mach's principle (general relativity) and entanglement (quantum theory) cannot be understood as a quantum gravity manifestation.

%%%%%%%%%%%%%%%%%%%%%%%%%%%%%%%%%%%%%%%%%%%%%%%%%%%%%%%%%%%%%%%%%%%%%%%%%%%%%%%%%%%%%%%%%%%%%%%%%%%%%%%%%%%%%%%%%%%%%%%%%%%%%%%%%%%%%%%%%%%%%%%%%%%%%%%%%%%%%%%%%%%%%%%%%%%%%%%%%%%%%%%%%%%%%%%%%%%%%%%%%%%%%%%%%%%%%%%%%%%%%%%%%%%%%%%%%%%%%%%%%%%%%%%%%%%%%%%%%%%%%%%%%%%%%%%%%%%%%%%%%%%%%%%%%%%%%%%%%%%%%%%%%%%%%%%%%%%%%%%%%%%%%%%%%%%%%%%%%%%%%%%%%%%%%%%%%%%%%%%%%%%%%%%%%

%%%%%%%%%%%%%%%%%%%%%%%%%%%%%%%%%%%%%%%%%%%%%%%%%%%%%%%%%%%%%%%%%%%%%%%%%%%%%%%%%%%%%%%%%%%%%%%%%%%%%%%%%%%%%%%%%%%%%%%%%%%%%%%%%%%%%%%%%%%%%%%%%%%%%%%%%%%%%%%%%%%%%%%%%%%%%%%%%%%%%%%%%%%%%%%%%%%%%%%%%%%%%%%%%%%%%%%%%%%%%%%%%%%%%%%%%%%%%%%%%%%%%%%%%%%%%%%%%%%%%%%%%%%%%%%%%%%%%%%%%%%%%%%%%%%%%%%%%%%%%%%%%%%%%%%%%%%%%%%%%%%%%%%%%%%%%%%%%%%%%%%%%%%%%%%%%%%%%%%%%%%%%%%%%

\section{Combining MOND theory and Machian model}
\renewcommand{\theequation}{8.\arabic{equation}}
\setcounter{equation}{0}

\noindent We will begin by writing the universal constant of the speed of light as \cite{gogberashvili}
\begin{eqnarray}
\label{c2}
c^2=\frac{2 M_U G}{R}\,\,.
\end{eqnarray}
From Eqs. (\ref{amond}) and (\ref{c2}) we can establish that

\begin{eqnarray}
\label{ac2}
 a=\frac{c^2}{2 R}\,\,,
\end{eqnarray}
where $a$ is the gravitational acceleration at the radius of the universe, $R$. Then, using Eqs. (\ref{amond}) and ({\ref{ac2}) again we can derive an expression for the speed of light compatible with MOND's theory

\begin{eqnarray}
\label{c2m}
 c^2=2 \sqrt{G M_U a_0}\,\,.
\end{eqnarray}

We can generalize Eq. (\ref{n0}) that describes the bits inhomogeneity by introducing a $\beta$ parameter in the following manner

\begin{eqnarray}
\label{n0b}
 \frac{N_0}{N_b}=1-{\(\frac{T}{T_c}\)}^\beta\,\,.
\end{eqnarray}
In the language of critical phenomena, the $\beta$ parameter can be identified as a critical exponent \cite{Huang}.
Following the same procedure of (\ref{ebec}) we have

\begin{eqnarray}
\label{ebecn}
M c^2 &=& \frac{1}{2} (N_b -N_0) k_B T \nonumber\\\nonumber\\ 
&=& \frac{1}{2}\, N_b \(\frac{T}{T_c}\)^\beta \; k_B T\,\,.
\end{eqnarray}

\noindent Then, combining Eqs. (\ref{bits}), (\ref{un}) and  (\ref{ebecn}) we have 

\begin{eqnarray}
\label{amondb}
 a=\(\frac{G M}{R^2} \, a_0^\beta\)^{\frac{1}{\beta+1}}\,\,.
\end{eqnarray}
Using Eq. (\ref{ac2})  we can find an expression for the speed of light compatible with Eq. (\ref{amondb})

\begin{eqnarray}
\label{c2b}
 c^2=2 R^{\frac{\beta-1}{\beta+1}}\; {(G M)}^{\frac{1}{\beta+1}}\; a_0^{\frac{\beta}{\beta+1}}\,\,.
\end{eqnarray}
It is important to note that Eq. (\ref{c2b}) is a generalization of the speed of light in the Machian model of the universe. As a consequence, for $\beta=0$ and $\beta=1$, we obtain Eqs. (\ref{c2}) and (\ref{c2m}) respectively. A very interesting result can be seen from Eq. (\ref{c2b}) if we use approximate values $R\approx 10^{25}\,m$, $G\approx 10^{-11} \,m^3 \,kg^{-1} \,s^{-2},\; M_u\approx 10^{53}\, kg$, $a_0\approx 10^{-10} \,m \,s^{-2}$ and substitute all of them in Eq. (\ref{c2b}). So we will have that

\begin{eqnarray}
\label{5.8}
 c^2=2 (10^{25}m)^{\frac{\beta-1}{\beta+1}}\; {(10^{42}m^3 s^{-2})}^{\frac{1}{\beta+1}}  (10^{-10}\,m \,s^{-2})^{\frac{\beta}{\beta+1}}.
\end{eqnarray}

\ni In order to have a better visualization of the result, let us obtain the logarithm of equation (\ref{5.8}).  After an algebraic calculation, we have that
\bbe \label{5.9}
log\,c\,=\,\frac{log\,2 }{2}\,+\,\frac 12 \,\frac{1}{\beta +1}\,[15\beta\,+\,17].
\eee
When $\beta \longrightarrow 0$ we have
\bbe \label{5.10}
log\,c_{0} \,=\,8.6 \qquad \Longrightarrow \qquad c_0\,=\,{10}^{8.6}\,m/s,
\eee
and when $\beta \longrightarrow \infty$ we obtain
\bbe \label{5.10b}
log\,c_{\infty}\,=\,7.6 \qquad \Longrightarrow \qquad c_{\infty}\,=\,{10}^{7.6}\,m/s\,\,,
\eee
which show us a decrease in the speed of light as a consequence of a increase of nonhomogeneous parameter, $\beta$. In figure 1 we have constructed the graphic showing the $log\,c$ behavior as a function of $\beta$.

\begin{figure}[ih]
\includegraphics[scale=1.1,angle=0]{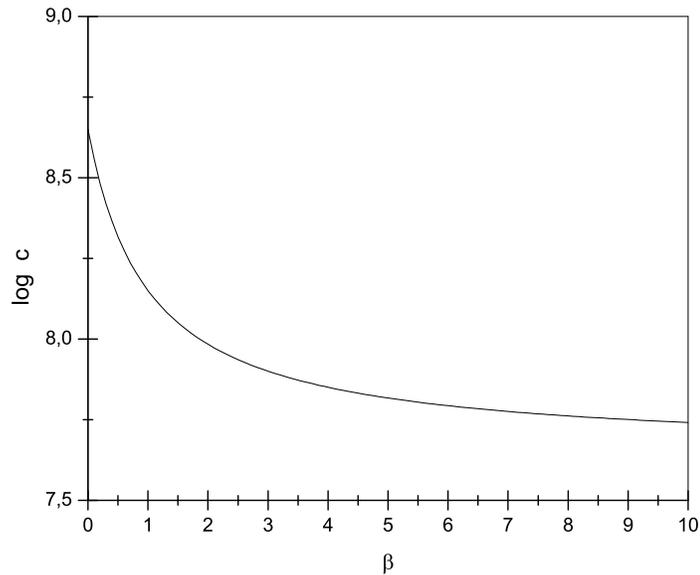}
\caption{Logarithm of speed of light versus $\beta$.}
\end{figure}

\section{Conclusions}
\label{Cs}

In this paper, by using the Tsallis thermostatistics formalism and the Verlinde approach of the entropic gravity, we have derived some physical quantities of the Machian model as a function of the $q$-nonextensive parameter. Besides, we have obtained a new value for $q$.

After that, we have constructed a relation between Mach's number of particles in the horizon and the number of bits of Verlinde's holographic screen.   Since this relation is very simple and direct we discussed about the possibility of a kind of mapping between both ideas.  Namely, the particle-particle long range interaction would be mapping in the bit-bit entanglement of the holographic screen.   since we have that the Mach's principle is an ingredient of general relativity and entanglement is a quantum property, we asked if the combination obtained here is not a quantum gravity manifestation.

In the context of the modified Newton dynamics, MOND, we have established an expression for the speed of light compatible with MOND's theory.  Furthermore, making use of critical phenomena and phase transition models, we have proposed another bits energy distribution as a function of temperature in order to introduce the notion of critical exponents. As a consequence, we have obtained a generalized gravitational acceleration. This important result allows us to derive in the Machian model a generalized expression for the speed of light in terms of a critical exponent. Then, we have verified that the speed of light, for distance equal to the radius of universe, decreases with the increase of the critical exponent.

\section{Acknowledgments}

\ni EMCA would like to thank CNPq (Conselho Nacional de Desenvolvimento Cient\' ifico e Tecnol\'ogico), Brazilian scientific support agency, for partial financial support.

\bigskip
\bigskip

%\newpage

\end{document}